\PassOptionsToPackage{unicode}{hyperref}
\PassOptionsToPackage{hyphens}{url}
\documentclass[
]{article}
\usepackage{amsmath,amssymb}
\usepackage{iftex}
\ifPDFTeX
  \usepackage[T1]{fontenc}
  \usepackage[utf8]{inputenc}
  \usepackage{textcomp} 
\else 
  \usepackage{unicode-math} 
  \defaultfontfeatures{Scale=MatchLowercase}
  \defaultfontfeatures[\rmfamily]{Ligatures=TeX,Scale=1}
\fi
\usepackage{lmodern}
\ifPDFTeX\else
\fi
\IfFileExists{upquote.sty}{\usepackage{upquote}}{}
\IfFileExists{microtype.sty}{
  \usepackage[]{microtype}
  \UseMicrotypeSet[protrusion]{basicmath} 
}{}
\usepackage{xcolor}
\usepackage[margin=1in]{geometry}
\usepackage{graphicx}
\makeatletter
\def\maxwidth{\ifdim\Gin@nat@width>\linewidth\linewidth\else\Gin@nat@width\fi}
\def\maxheight{\ifdim\Gin@nat@height>\textheight\textheight\else\Gin@nat@height\fi}
\makeatother
\setkeys{Gin}{width=\maxwidth,height=\maxheight,keepaspectratio}
\makeatletter
\def\fps@figure{htbp}
\makeatother
\NewDocumentCommand\citeproctext{}{}

\makeatletter
 \let\@cite@ofmt\@firstofone
 \def\@biblabel#1{}
 \def\@cite#1#2{{#1\if@tempswa , #2\fi}}
\makeatother
\newlength{\cslhangindent}
\newlength{\csllabelwidth}
\newenvironment{CSLReferences}[2] 
 {\begin{list}{}{%
  \setlength{\itemindent}{0pt}
  \setlength{\leftmargin}{0pt}
  \setlength{\parsep}{0pt}
  \ifodd #1
   \setlength{\leftmargin}{\cslhangindent}
   \setlength{\itemindent}{-1\cslhangindent}
  \fi
  \setlength{\itemsep}{#2\baselineskip}}}
 {\end{list}}
\usepackage{calc}

\usepackage{mathtools}
\DeclareMathAlphabet{\mathams}{U}{msb}{m}{n}

\usepackage{booktabs}
\usepackage{float}
\usepackage{booktabs}
\usepackage{longtable}
\usepackage{array}
\usepackage{multirow}
\usepackage{wrapfig}
\usepackage{float}
\usepackage{colortbl}
\usepackage{pdflscape}
\usepackage{tabu}
\usepackage{threeparttable}
\usepackage{threeparttablex}
\usepackage[normalem]{ulem}
\usepackage{makecell}
\usepackage{xcolor}
\ifLuaTeX
  \usepackage{selnolig}  
\fi
\usepackage{bookmark}
\IfFileExists{xurl.sty}{\usepackage{xurl}}{} 
\hypersetup{
  pdftitle={Fast fitting of phylogenetic mixed-effects models},
  pdfauthor={Bert van der Veen\^{}\{1\}; Robert B. O'Hara\^{}\{1\}},
  pdfkeywords={jsdm; gllvm; multivariate; fourth corner; phylogenetic
comparative methods; NNGP,},
  hidelinks,
  pdfcreator={LaTeX via pandoc}}

\title{Fast fitting of phylogenetic mixed-effects models}
\author{Bert van der Veen\(^{1}\) \and Robert B. O'Hara\(^{1}\)}
\date{\(^1\)Department of Mathematical Sciences, Norwegian University of
Science and Technology, Trondheim, Norway \newline}

\begin{document}
\maketitle
\begin{abstract}
\noindent Mixed-effects models are among the most commonly used
statistical methods for the exploration of multispecies data. In recent
years, also Joint Species Distribution Models and Generalized Linear
Latent Variale Models have gained in popularity when the goal is to
incorporate residual covariation between species that cannot be
explained due to measured environmental covariates. Few software
implementations of such models exist that can additionally incorporate
phylogenetic information, and those that exist tend to utilize Markov
chain Monte Carlo methods for estimation, so that model fitting takes a
long time. In this article we develop new methods for quickly and
flexibly fitting phylogenetic mixed-effects models, potentially
incorporating residual covariation between species using latent
variables, with the possibility to estimate the strength of phylogenetic
structuring in species responses per environmental covariate, and while
incorporating correlation between different covariate effects. By
combining Variational approximations, a sparse approximation to the
phylogenetic precision matrix, and parallel computation, phylogenetic
mixed-effects models can be fitted much more quickly than the current
state-of-the-art. Two simulation studies demonstrate that the proposed
combination of approximations is fast and enjoys high accuracy. We
explore sensitivty of the approximation to the ordering of species with
a real world dataset of wood-decaying fungi.
\end{abstract}

Word count: \ensuremath{1.1172\times 10^{4}}

\section{Introduction}\label{introduction}

The exploration of species co-occurrence patterns is at the core of
contemporary community ecology. Recent years have seen a push in methods
that better facilitate ecologists to understand which species co-occur,
and why they co-occur (Ives and Helmus 2011; Braga et al. 2018). Some
species co-occur because they thrive in similar environments, which may
be due to similar characteristics in terms of functional traits. Species
that share evolutionary history often share such physical
characteristics, and consequently hypotheses exist that describe the
co-occurrence of closely related species (Pagel 1999).

Data on ecological communities is increasingly analysed with
multivariate statistical models. Joint Species Distribution Models
(JSDMs, Pollock et al. 2014) are now frequently applied for the analysis
of binary data of multiple species. JSDMs are a type of Generalized
Linear Mixed-effects Model (GLMM, Brooks et al. 2017; Bates et al. 2015)
that incorporates residual covariation to better represent species'
co-occurrence patterns. Generalized Linear Latent Variable Models
(GLLVMs, Warton et al. 2015) are a modeling framework that leverages
dimension reduction methods to more efficiently implement such
multispecies models. GLLVMs have also been presented as a model-based
method for ordination (Hui et al. 2015; van der Veen et al. 2023), and
have been at the forefront of computational advances for quickly fitting
statistical models with random effects (Niku et al. 2019; Korhonen et
al. 2021; Kidzinski et al. 2022).

Multivariate statistical models tend to be challenging to fit with
species that do not have many observations. Such species may be referred
to as ``rare'' by ecologists, but we here more generally refer to
species that have few observations, such as due to insufficient
sampling. In order to ameliorate such issues with few data points, and
to incorporate the relatedness of species into the analysis of species
co-occurrence patterns, JSDMs have been developed that phylogenetically
structure species responses to the environment. In such models, species
that are more closely related can be predicted to co-occur (Ovaskainen
et al. 2017). This has the benefit of ``sharing'' information across
related species, so that species may more accurately be placed in the
environment. In fact, with such models it is possible to predict the
environmental preferences of extinct species for which no response
observations are available at all, by utilizing phylogenetic
information.

Three notable software implementations for such phylogenetic
mixed-effects models are the \texttt{Hmsc} (Tikhonov et al. 2024),
\texttt{MCMCglmm} (Hadfield 2010), and \texttt{phyr} (Ives et al. 2020)
\texttt{R}-packages. Unfortunately, these software implementations are
either slow because they use Markov chain Monte Carlo for estimation
(\texttt{Hmsc} and \texttt{MCMCglmm}) or are limited in their
functionality otherwise (e.g., \texttt{phyr} supports few datatypes).
Since phylogenetic information and community data of species is becoming
more readily available thanks to various openscience initiatives
{[}Hinchliff et al. (2015);{]}, this motivates development of better,
and faster, software implementations of community models that
incorporate phylogenetic information.

In this article, we present a new statistical approach for fast fitting
of phylogenetic mixed-effects models for community ecological data as
part of the \texttt{gllvm} \texttt{R}-package (Niku et al. 2023).
Phylogenetic information can facilitate better estimation of species
responses to the environment, while latent variables can potentially be
included to account for residual covariation. The proposed modeling
framework can alternatively be used for when response data are
functional traits, or data on individuals (with a pedigree instead of a
phylogeny as in \texttt{MCMCglmm}), but we choose to focus on community
data in this article. The software implementation supports a range of
datatypes commonly found in community ecology, such as for counts of
individuals (Poisson and negative-binomial, potentially zero-inflated),
cover classes (ordinal), percentage cover (beta models potentially with
zeros or ones, Korhonen et al. 2024), and biomass (Tweedie models,
Korhonen et al. 2021; Niku et al. 2017). Niku et al. (2021) implemented
species-specific random effects as part of the fourth corner latent
variable model. We extend their work by; 1) phylogenetically structuring
the random effects, 2) for which we provide functionality to determine
strength of the phylogenetic signal per covariate, and 3) facilitate
specification of correlation between the random effects for different
covariates using the user-friendly \texttt{lme4} formula interface
(Bates et al. 2015, 2025)

Using a combination of methodological developments we can fit the models
much more quickly than existing software implementations for
phylogenetic mixed-effects models. Variational Approximations (VA,
Ormerod and Wand 2010) helps to retrieve a closed form approximation to
the likelihood, which we combine with a matrix normal covariance
structure for the variational distribution, as well as a reduced-rank
approximation to further speed it up. A sparse approximation to the
phylogenetic precision matrix, in the form of a band pattern (Wist and
Rue 2006) or with Nearest Neighbour Gaussian processes (NNGPs, Finley et
al. 2019) facilitates even quicker computation, i.e., by fitting the
models using Gaussian Markov Random fields. Recently, Matsuba et al.
(2024) also employed NNGPs to quickly and accurately retrieve the
inverse of the phylogenetic covariance matrix, although in a Bayesian
context and following for a different model formulation. Finally, we fit
the models in parallel via the \texttt{TMB} \texttt{R}-package
(Kristensen et al. 2015) to further reduce computation time.

With extensive simulations we show that the proposed combination of four
approximations still allows us to accurately retrieve the model
parameters, and that it tends to be even faster than fitting the models
with the Laplace approximation (e.g., as used in the \texttt{phyr}
\texttt{R}-package). The approximation error for the phylogenetic
precision matrix is sensitive to row-column permutation (Guinness 2018),
an issue that we explore using real data on wood-decaying fungi from
Abrego et al. (2022a).

\section{Model formulation}\label{model-formulation}

In the following sections we use superscripts to distinguish different
quantities, subscripts for indices, bold capitalized characters for
matrices, bold lowercase characters for vectors, and ordinary characters
to denote scalars. In this section we present and explain the
phylogenetic mixed-effects model, and in the next section we discuss
estimation of the parameters.

Let \(\textbf{Y}\) represent a multivariate dataset with observations
\(y_{ij}\), where the rows \(i = 1,\ldots, n\) represent the sites at
which an ecological community has been surveyed, and the columns
\(j = 1, \ldots, m\) represent the species found. Let
\(\textbf{X} = \begin{pmatrix}\textbf{x}_1, \textbf{x}_2, \ldots, \textbf{x}_k\end{pmatrix}\)
denote a set of \(k = 1 ,\ldots, p\) covariates, where
\(\textbf{x}_i = \begin{pmatrix}x_{i1}, x_{i2}, \ldots x_{nk} \end{pmatrix}^\top\)
with an associated \(p \times m\) coefficient matrix \(\textbf{B}\) and
entries \(\beta_{jk}\) for the environment that has been observed at
each site, usually including an intercept column. Covariates can be
categorical to represent species-specific intercepts under different
conditions, or numerical to represent species-specific responses to the
environment, i.e., as slope parameters. As in generalised linear models
(GLMs) a link function \(\text{g}(\cdot)\) connects the model to the
conditional mean of the distribution assumed for the responses, so that
the model generically is

\begin{equation}
\text{g}\{\mathams{E}(\textbf{Y}\vert\textbf{B})\} = \textbf{X}\textbf{B}.
\label{one}
\end{equation}

This model can additionally be extended with functional traits as in the
fourth corner latent variable model (Niku et al. 2021), by
hierarchically modeling the species responses

\begin{equation}
\textbf{B} = \textbf{1}^\top\boldsymbol{\beta}^x + \textbf{T}\textbf{B}^{tx} + \textbf{B}^\epsilon,
\nonumber
\end{equation}

\noindent where \(\textbf{1}\) is a \(m\)-sized vector of ones,
\(\boldsymbol{\beta}^x = \begin{pmatrix}\beta^x_1, \beta_2, \ldots, \beta^x_p\end{pmatrix}^\top\)
is a \(p\)-sized vector of community mean responses to the covariates,
\(\textbf{T}\) is a \(m \times t\) matrix of species covariates (most
commonly functional traits), \(\textbf{B}^{tx}\) is a \(t \times p\)
matrix of interaction coefficients with columns
\(\boldsymbol{\beta}^{tx}_k\), and \(\textbf{B}^\epsilon\) is a matrix
of \(p \times m\) species-specific random effects for the covariates
that model species' deviations from the community mean response with
entries \(\beta^\epsilon_{jk}\). The \(\textbf{B}\) with coefficients
can be thought of as latent traits, which are modeled with species
covariates. Excluding the species covariates thus simplifies the model
to only include ``residual'' information.

Let \(\textbf{C}\) denote the phylogenetic correlation matrix of
dimension \(m \times m\), and
\(\boldsymbol{\mu}_k = \boldsymbol{\beta}^x_k + \textbf{T}\boldsymbol{\beta}^{tx}_k\)
the community mean response to the \(k^{th}\) covariate. To
phylogenetically structure the random effects we could assume a model
with a single set of random effects
\(\textbf{b}^a_{k} \sim \mathcal{N}(\boldsymbol{\mu}_k, \gamma_k^2\textbf{C})\),
so that the variance of the random effect \(\gamma_k^2\) represents the
rate of evolution. However, in the absence of a phylogenetic signal this
model is quite restrictive, so phylogenetic random effects models are
usually formulated with a second set of random effects that are
independent for all species, \(\textbf{b}^2_{k}\) with zero mean and
variance \(\delta_k^2\). Since \(\textbf{b}^a_k\) and \(\textbf{b}^b_k\)
are independent, it follows that
\(\boldsymbol{\beta}_k = \textbf{b}^a_k + \textbf{b}^b_k \sim \mathcal{N}(\boldsymbol{\mu}_k, \boldsymbol{\Sigma}_k)\)
where
\(\boldsymbol{\Sigma}_k = \sigma^2_k\{\textbf{C}\rho_k + (1-\rho_k)\textbf{I}^m\}\)
and \(\rho_k = \gamma_k^2/(\gamma^2_k+\delta_k^2)\) is the phylogenetic
signal parameter per covariate (equivalent to Pagel's \(\lambda\), Pagel
1999; Pearse, Davies, and Wolkovich 2023), while
\(\sigma^2_k = \gamma^2_k + \delta^2_k\) is the variance of the sum of
the two terms. Consequently, the variances of the original two terms can
be retrieved from this alternative model formulation by nothing that
\(\gamma^2_k = \rho_k\sigma_k^2\) and
\(\delta^2_k = (1-\rho_k)\sigma^2_k\). Equivalently,
\(\boldsymbol{\beta}_k^\epsilon \sim \mathcal{N}(\textbf{0}, \boldsymbol{\Sigma}_k)\).
The latter formulation has the benefit of only including one set of
random effects that needs to be integrated out of the likelihood (which
is usually computationally intensive for many random effects), but the
downside that it requires explicit inversion of the covariance matrix
for species at each iteration during model fitting (which tends to be
computationally intensive for many species).

Covariances between the random effects of different covariates
\(\boldsymbol{\Sigma}^r\), i.e., for the rows of \(\textbf{B}^\epsilon\)
can be introduced by instead defining the distribution of the random
effects over all species and covariates simultaneously. Let
\(\textbf{L}(\boldsymbol{\Sigma}_k)\) denote the lower cholesky factor
of the covariance matrix for the \(k^{\text{th}}\) covariate effect, so
that instead we parameterize the covariance over the random effects for
all species and covariates as

\begin{equation}
\text{vec}(\textbf{B}^{\epsilon\top}) \sim \mathcal{N}\{\textbf{0},\textbf{L}(\boldsymbol{\Sigma}^r \otimes \textbf{I}^m)\textbf{L}^\top\},
\label{two}
\end{equation}

\noindent where \(\textbf{L}\) is a block-diagonal matrix with
\(\textbf{L}(\boldsymbol{\Sigma}_1), \dots, \textbf{L}(\boldsymbol{\Sigma}_p)\)
as block matrices. In this formulation, \(\boldsymbol{\Sigma}^r\)
controls correlation between covariate effects of the same species, and
together with the phylogenetic signal parameters also the correlation
between covariate effects of different species. These off-diagonal
blocks in the covariance matrix are by extension formulated as a
lower-upper (LU) decomposition and take the form
\(\textbf{L}(\boldsymbol{\Sigma}_k)\textbf{L}(\boldsymbol{\Sigma}_l)^\top\Sigma_{kl}\),
for covariates \(k,l = 1\ldots p\). The lower triangular matrix of this
LU decomposition is
\(\textbf{L}(\boldsymbol{\Sigma}_k)\text{Diag}[\text{diag}\{\textbf{L}(\boldsymbol{\Sigma}_k)\}^{-1}\),
where \(\text{Diag}(\cdot)\) constructs a diagonal matrix from a vector
and \(\text{diag}(\cdot)\) extracts the diagonal of a matrix. The
associated upper triangular matrix is then
\(\text{Diag}[\text{diag}\{\textbf{L}(\boldsymbol{\Sigma}_k)\}]\textbf{L}(\boldsymbol{\Sigma}_l)^\top\Sigma_{kl}\).
If instead we assume \(\rho_k = \rho\), i.e., that the phylogenetic
signal is the same for all covariates, equation \eqref{two} reduces to
equation (4) of Ovaskainen et al. (2017) or equation (11.1-2) of de
Villemereuil and Nakagawa (2014), i.e.,
\(\text{vec}(\textbf{B}^\epsilon) \sim \mathcal{N}(\textbf{0},\boldsymbol{\Sigma} \otimes \boldsymbol{\Sigma}^r)\).

Consequently, the matrix of species associations at sites \(i\) and
\(i2\) due to the phylogenetic mixed-effects model with multiple signal
parameters is:

\begin{equation}
\boldsymbol{\Sigma}_{spp,i,i2} = 
(\textbf{x}_i^\top \otimes \textbf{I}^m)\textbf{L}(\boldsymbol{\Sigma}^r \otimes \textbf{I}^m)\textbf{L}^\top(\textbf{x}_{i2} \otimes \textbf{I}^m).
\nonumber
\end{equation}

The model in equation \eqref{one} can be extended in various ways, for
example by incorporating unconstrained latent variables as in Niku et
al. (2021), or constrained or informed latent variables (van der Veen et
al. 2023), potentially with a quadratic response model (van der Veen et
al. 2021), with random site effects to account for pseudoreplication (as
included in the following case study) or spatial autocorrelation, or
with additional fixed effects, but we have omitted those here for
brevity.

Similarly, in its present form the model only incorporates positive
species associations. It is technically possible to adjust the model to
additionally incorporate negative species assocations, but we consider
such an implementation as a potential future avenue for research (see
appendix S1 for more information).

\section{Parameter estimation}\label{parameter-estimation}

In this section we elaborate on the four approximations that are
implemented to quickly fit the phylogenetic random effect models. We
evaluate the accuracy of these stacked approximations using simulation
studies in the following section.

For brevity, formulating the model in terms of a single set of random
effects with covariance matrix \(\boldsymbol{\Sigma}\) given in equation
\eqref{two}. With \(f(y_{ij} \vert \boldsymbol{\beta}^\epsilon_j)\) a
response distribution such as a Tweedie or negative binomial
distribution, and \(\Theta\) a vector incorporating of all other
parameters in the model (such as scale parameters), we have the
following marginal log-likelihood function:

\begin{equation}
\mathcal{L}(\boldsymbol{\Theta}) =\sum \limits_{i=1}^n \log\biggl[\int \prod \limits_{j=1}^m f\biggl\{y_{ij} \vert \boldsymbol{\beta}^\epsilon_j\biggr\}\mathcal{N}\biggl\{\text{vec}\biggl(\textbf{B}^\epsilon\biggr); \boldsymbol{\Sigma}\biggr\} d\textbf{B}^\epsilon\biggr].
\label{three}
\end{equation}

\subsection{Variational
Approximations}\label{variational-approximations}

To efficiently retrieve a closed-form approximation to the
log-likelihood in equation \eqref{three}, we use Variational
Approximations (VA, Hui et al. 2017). VA introduces hyperparameters
(``variational parameters'') that need to be optimised in order to
improve the approximation (tightening the lower bound). With many
species and random effects this can significantly slow down model
fitting. Other methods for fitting generalized mixed-effects models,
such as the Laplace approximation, suffer from similar issues with high
dimensions. The Laplace approximation requires evaluating the matrix of
second derivatives of the joint likelihood for the random effects and
the data, which also significantly slows down for a large number of
(correlated) random effects. VA has the benefit of allowing for a second
approximation by introducing a more simple structure for the VA
covariance matrix, so that the size of the optimisation problem is
reduced significantly for quicker model fitting.

Various VA covariance structures are possible to formulate, but some
require stronger assumptions (such as independence of species or random
effects), or have a considerably higher number of variational parameters
(e.g., as in an unstructured covariance matrix), so that they require
significantly longer computational times. In one of the most reduced
scenarios, we assume the variational distribution for the random effects
to follow a matrix normal distribution, i.e.,
\(q(\textbf{B}^\epsilon) = \mathcal{MN}(\textbf{a},\textbf{A}^r, \textbf{A}^m)\)
with \(\textbf{a}\) a \(p \times m\) matrix of variational mean
parameters, \(\textbf{A}^r\) a \(p \times p\) VA covariance matrix, and
\(\textbf{A}^m\) a \(m \times m\) VA covariance matrix. Some different
formulations are provided in the supplementary information (appendix S2)
and implemented in the \texttt{gllvm} \texttt{R}-package. For the
remainder of this article, we limit ourselves to application of the
matrix normal case, with unstructured or diagonal \(\textbf{A}^r\). We
show in the simulations below that the choice of a matrix normal VA
distribution still enjoys high accuracy, despite its relative
inflexibility, even with diagonal \(\textbf{A}^r\).

With this choice of a matrix normal covariance structure, VA can still
be slow due to the potentially large number of variational parameters in
\(\textbf{A}^m\). The number of variational parameters in this matrix
grows quadratically in the number of species. To ensure better scaling,
and quick computation also with many species, we employ a third
approximation by representing the off-diagonal entries of
\(\textbf{A}^m\) in a reduced-rank fashion so that \(d << m\), i.e.,
\(\textbf{A}^m = \textbf{A}^d\textbf{A}^{d\top}+ \textbf{A}^{m-d}\)
where \(\textbf{A}^d\) is a \(m \times d\) lower triangular matrix with
positive diagonal entries, and \(\textbf{A}^{m-d}\) is a \(m \times m\)
diagonal matrix with the first \(d\) diagonal entries equal to zero and
the remaining entries positive, ensuring that \(\textbf{A}^m\) is of
full rank. Such a choice for reducing the number of variational
parameters in the likelihood is optimal for computation speed, to reduce
memory use during model compilation, and to reduce the size of the
Hessian matrix that needs to be inverted for parameter estimation and
for post-hoc estimating (approximate) statistical uncertainties. This
does introduce the additional problem of having to select the rank of
the variational covariance matrix, but in the simulation studies that
follow we show that high accuracy can be achieved even when using only a
single dimension.

\subsection{Sparse precision matrix}\label{sparse-precision-matrix}

The final remaining computational bottleneck involves inversion of the
covariance matrix for the random effects. Since the covariance matrix of
the covariate effects \(\boldsymbol{\Sigma}^r\) is usually relatively
small, its inverse can be evaluated explicitly without excessive
computational cost, and its determinant straightforwardly retrieved by
formulating it following a log-cholesky parameterization (Pinheiro and
Bates 1996). In contrast, the number of species in multivariate
ecological data is often relatively large, so that
\(\boldsymbol{\Sigma}^{-1}_k\) is expensive to calculate. With a single
phylogenetic signal parameter over all covariate effects, the form of
the covariance function for \(\boldsymbol{\Sigma}_k\) allows inversion
by repeatedly applying the Sherman-Morrison formula for computationally
cheap rank-one updates of the form
\(\textbf{C}^{-1} = \textbf{C}^{-1} - \{1-\rho\}/\{1+(1-\rho)\textbf{C}_j\textbf{C}_j^\top\}\),
but further improvement is warranted.

Let \(A_j\) a set of indices (species) used to condition a random effect
on. Consequently, the joint distribution for the random effect over all
species can be written as
\(p(\boldsymbol{\beta}^\epsilon_k) = \prod \limits^{j=1} p(\beta^\epsilon_{jk} \vert \{\beta^\epsilon_{1k}, \beta^\epsilon_{2k}, \ldots, \beta^\epsilon_{j-1k}\} :A_j)\),
and while momentarily ignoring potential correlations between random
effects, each conditional distribution can be written as
\(p(\beta^\epsilon_{jk} \vert \{\beta^\epsilon_{1k}, \beta^\epsilon_{2l}, \ldots, \beta^\epsilon_{j-1k}\} :A_j) = \mathcal{N}(\boldsymbol{\Sigma}_{kjA_j}\boldsymbol{\Sigma}_{kA_jA_j}^{-1}\boldsymbol{\beta}^\epsilon_{A_j k}, \boldsymbol{\Sigma}_{kjj} - \boldsymbol{\Sigma}_{kjA_j}\boldsymbol{\Sigma}_{kA_jA_j}^{-1}\boldsymbol{\Sigma}_{kA_jj})\).

Wist and Rue (2006) describe various methods for specifying Gaussian
Markov Random Fields with a sparse precision matrix, such as by
thresholding or by a predefined sparsity pattern (such as a band
pattern). Here, we mostly focus on adopting NNGPs as our fourth and
final approximation, but compare with approximating the precision matrix
by a band pattern in the real data example. Unlike regular matrix
inversion, that tends to be inherently sequential (such as repeated
application of the Sherman-Morrison formula), NNGPs facilitate parallel
computation well because the calculation can be performed for all random
effects separately. In NNGPs we construct the (approximate) inverse of
the cholesky factor of the covariance matrix, which makes for an
efficient implementation of equation \eqref{two}, and combines nicely
with the likelihood in equation \eqref{three}.

NNGPs are usually applied in a spatial (and Bayesian) context (Finley et
al. 2019), truncating the set of indices \(A_j\) for spatial locations
based on geographical proximity. For a phylogenetic random effect we
instead condition each species' random effect on the species closest in
evolutionary space. Distance between species in evolutionary space is
readily accessible from the phylogeny, and calculated as the total
shared branch length on the phylogeny. It is this truncation for the
conditional distributions of the random effects that implies a sparse
precision matrix for the joint distribution of the random effects. Note,
NNGPs are order-sensitive, so that the approximation can suffer from a
poor ordering of species. However, similarly to Guinness (2018), we
choose to see this as a feature of the methodology, so that the
approximation can be improved by permuting the columns of the response
data. We further explore this issue of order sensitivity in the real
data example.

\section{Simulation studies}\label{simulation-studies}

To evaluate the accuracy of the proposed combination of approximations,
we performed various simulation studies, but for brevity we assumed
\(\rho_k = \rho\), i.e., a single phylogenetic signal parameter for all
covariates. We performed simulations separately to evaluate the accuracy
of the low-rank VA approximation, and of the NNGP approximation. To do
so, we simulated datasets following equation \eqref{one}, without
functional traits, and while assuming that the phylogenetic signal
parameter is the same for all covariates. We focus on estimation
accuracy of the phylogenetic signal, as it conveniently provides a
single measure to assess, but also because we deem it the most likely
quantity in the model to be affected by changes in the approximation.

For the first simulation study, we kept the number of sites fixed at
\(n = 100\) and the number of species at \(m = 200\). We simulated 250
datasets while varying the rank of the VA covariance matrix
(\(d \in \{1, 3, 5, 10, 25, 100 , 200\}\)). Instead of the NNGP, in this
first simulation study we directly calculated the inverse of the
phylogenetic covariance, in order to avoid introducing potential error
due to the NNGP and so that we may isolate the error solely due to the
matrix normal VA approximation with reduced rank, and unstructured or
diagonal \(\textbf{A}^r\). For comparison we additionally fitted the
model with the Laplace approximation as automatically applied by
Template Model Builder (TMB, Kristensen et al. 2015). We re-simulated
the phylogeny for each simulation using the \texttt{ape}
\texttt{R}-package (Paradis and Schliep 2019). We simulated five
covariates from \(\text{Uniform}(-1, 1)\), the covariance matrix for the
random effects from \(W(\frac{1}{5}\textbf{I}^p, p)\), and kept
\(\rho = 0.5\) fixed so that the median absolute error is at best zero
and at most 0.5. As the total number of simulations is high, and the
models take long to fit when the rank of the VA approximation is large,
we only simulated binary datasets. Binary datasets contain least
information of the datatypes that are accommodated by the \texttt{gllvm}
\texttt{R}-package, so that these simulations should provide a reliable
upper bound to the accuracy of the approximation. The results of the
simulations are summarized in Figure 1.

\begin{figure}
\centering
\includegraphics{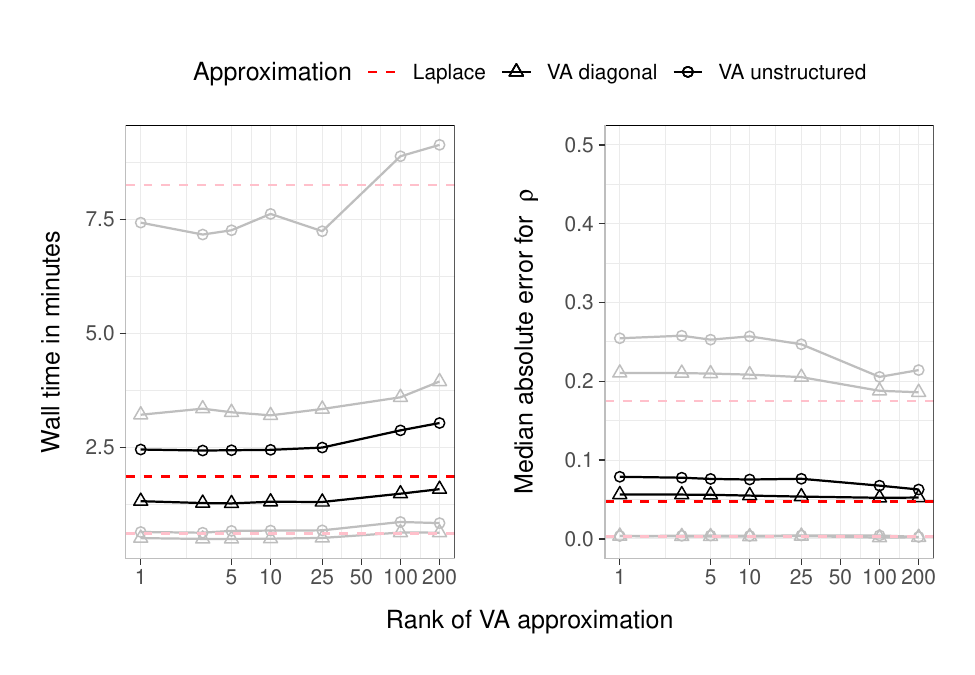}
\caption{Results of the first simulation study for evaluating the
accuracy of the low-rank VA approximation. The left panel displays
increase of fitting time with the rank (1, 3, 5, 10, 25, 100, 200) of
the VA approximation, with solid black lines indicating models fitted
with the VA approximation and with unstructured \(\textbf{A}^r\)
(circles) or diagonal \(\textbf{A}^r\) (triangles), while the dashed red
line indicates the computation time for the models fitted by Laplace
approximation. The faded black lines represent the 2.5\% and 97.5\%
percentiles for the VA approximation and the faded dashed red lines for
the Laplace approximation. The right panel shows the median absolute
error for the phylogenetic signal parameter, which at most can be 0.5.}
\end{figure}

The first simulation study confirms that the computation time of the
models rapidly increase with the selected rank of \(\textbf{A}^m\) in
the VA approximation, while being competitive to the Laplace
approximation implemented with TMB (indicated by the red dashed line).
As seen in the left panel of Figure 1, the rank of the VA approximation
has a large impact of the computation time, but fortunately little
impact on the estimation accuracy of the phylogenetic signal parameter,
which remained nearly stable across various ranks of the VA
approximation and was indeed lowest for the full rank approximation.
Models fitted with an unstructured \(\textbf{A}^r\) in the VA
approximation exhibited longer fitting times, so that it took longer to
fit models than with the Laplace approximation. Models fitted with a
diagonal \(\textbf{A}^r\) fitted more quickly than with the Laplace
approximation, while also providing more accurate estimates for the
phylogenetic signal than with an unstructured \(\textbf{A}^r\).

In the second simulation study we kept the rank of the VA approximation
fixed to one and \(\textbf{A}^r\) unstructured, and fixed the number of
sites to \(n = 100\), but instead varied the number of species
(\(m \in \{50, 100, 250, 500\}\)) and the number of nearest neighbors
used (\(nn \in \{1, 3, 5, 7, 9, 11, 13, 15\}\)) to approximate the
inverse of the phylogenetic covariance matrix. For each combination we
simulated 250 datasets, for either binary data or counts. The results
are presented in Figure 2.

\begin{figure}
\centering
\includegraphics{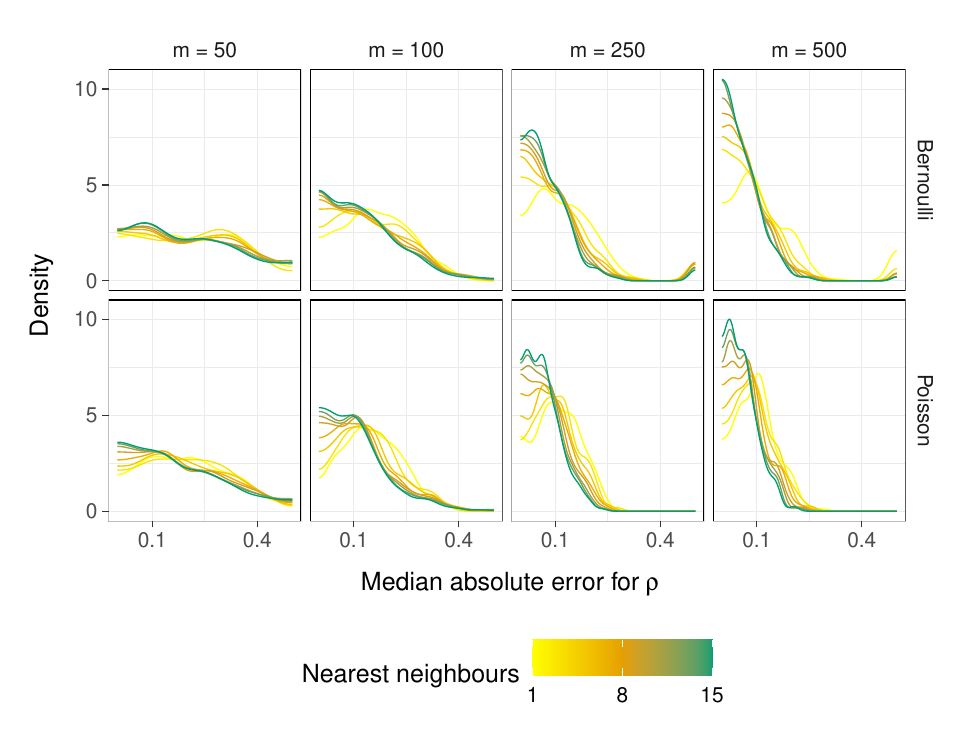}
\caption{Results for the second simulation study to test the ability of
the models to accurately estimate the phylogenetic signal parameter
using the low-rank VA and NNGP approximations for different numbers of
species \(m\) in the data.}
\end{figure}

The results in Figure 2 represent the compounded approximation error due
to both the low-rank VA approximation and the NNGP. The number of
species has a larger impact on the estimation accuracy of the
phylogenetic signal than the NNGP approximation. As the number of
species increases, the estimation accuracy of the phylogenetic signal
improves, with \(m = 50\) potentially requiring a much larger number of
nearest neighbouurs for accurate estimation, but with \(m>100\)
providing accurate estimates under most scenarios. As expected, the
accuracy was higher with more nearest neighbours, but a relatively low
(e.g., 10) number of nearest neighbors seems sufficient for reasonably
accurate estimates of the phylogenetic signal, even for large numbers of
species in the data. Though not presented here, application of NNGPs
reduced computation time significantly relative to the models fitted for
the first simulation study.

\section{Case study}\label{case-study}

We demonstrate the new method for fitting phylogenetic random effect
models on data collected by Abrego et al. (2022a). The dataset contains
the binary responses of 320 wood-inhabiting fungi surveyed on 1809 logs
across 53 European beech (\emph{Fagus sylvatica}) forest sites in
different European countries that were grouped into eight regions, so
that there were in total \(n = 1666\) rows in the data. Species with
four or less presences were removed before analysis in the original
article. Abrego et al. (2022a) fitted JSDMs with the \texttt{Hmsc}
\texttt{R}-package with as main goal to determine if traits and
phylogenetic relationships structure species communities differently at
different spatial scales. To that end, their analyses included
environmental variables both at the log- and site-scales such as:
diameter at breast height of logs (DBH), decay stage of the logs (on a
scale of one to five), an index for connectivity (at the 10km scale),
area of forest patches (in hectares), annual temperature range (degrees
Celsius) and annual precipitation, and naturally a phylogeny.

The models fitted by Abrego et al. (2022a) additionally incorporated six
latent variables at the region-level, and although the \texttt{gllvm}
\texttt{R}-package facilitates including such latent variables at the
group-level, we have here chosen to omit the latent variables for
brevity. In the \texttt{Hmsc} \texttt{R}-package each model took about a
week to 10 days to run (N. Abrego, pers. comm., 1st March 2024). Results
for the same four models as in are included in appendix S3. It takes
about six to twenty minutes to fit a model with covariate-specific
phylogenetic signal parameters (on a Dell Latitude 7490 using 7 CPU,
with \(nn = 15\) neighbours, and with unstructured \(\textbf{A}^r\)),
and \textbf{two to four minutes} for models while assuming
\(\rho_k = \rho\), i.e., a single phylogenetic parameter for all
covariate effects. Fitting the models with diagonal \(\textbf{A}^r\) can
further reduce computation time, to (e.g.) 1.5 minute (corresponding to
the two minutes for a model without traits). These models incorporated
nested random intercepts for the study region and reserves, and
including all correlation parameters between the covariate effects.

As previously noted, the quality of the NNGP approximation tends to be
sensitive to the order of species. This has previously been noted by
Guinness (2018) in a spatial context, who also explored orderings for
improving the NNGP approximation. Similarly to finding permutation
methods for reducing fill-in of the cholesky factor of a sparse matrix
(Wist and Rue 2006), identifying an optimal ordering for NNGPs is an
NP-complete problem. Consequently, heuristic methods are essential for
determining an ordering that improves the NNGP approximation.

The approximation error for the phylogenetic precision matrix will have
to be assessed on a case-by-case basis, but can be explored before model
fitting, because the approximation error for \(\boldsymbol{\Sigma}_k\)
is clearly related to that of \(\textbf{C}\). However, the former is not
available before model fitting, whereas the approximation to
\(\textbf{C}^{-1}\) is. We define the approximation error as
\(\|\widetilde{\textbf{C}^{-1}}\textbf{C}-\textbf{I}^m\|_F\), where
\(\widetilde{\textbf{C}^{-1}}\) denotes the approximation to
\(\textbf{C}^{-1}\).

Rather than extensively searching for potentially suitable orderings, we
instead focus on exploring the approximation error due to six orderings
that are easily accessible to ecologists: 1) the order of species in the
phylogeny; because species with large covariances are naturally located
closely together in the phylogeny, 2) alphabetically; as a baseline and
because it is how data is often provided in practice, 3) the sum of
pairwise distances of species, 4) the distance of species to the root of
the phylogeny, 5) by the first eigenvector of \(\textbf{C}\) (as in
Matsuba et al. 2024), 6) by the sum of squared covariances. We
additionally explore the approximation error using a band pattern (as in
Wist and Rue 2006), ignoring the distance between species in the
phylogeny. The results are visualized in Figure 3.

\begin{figure}
\centering
\includegraphics{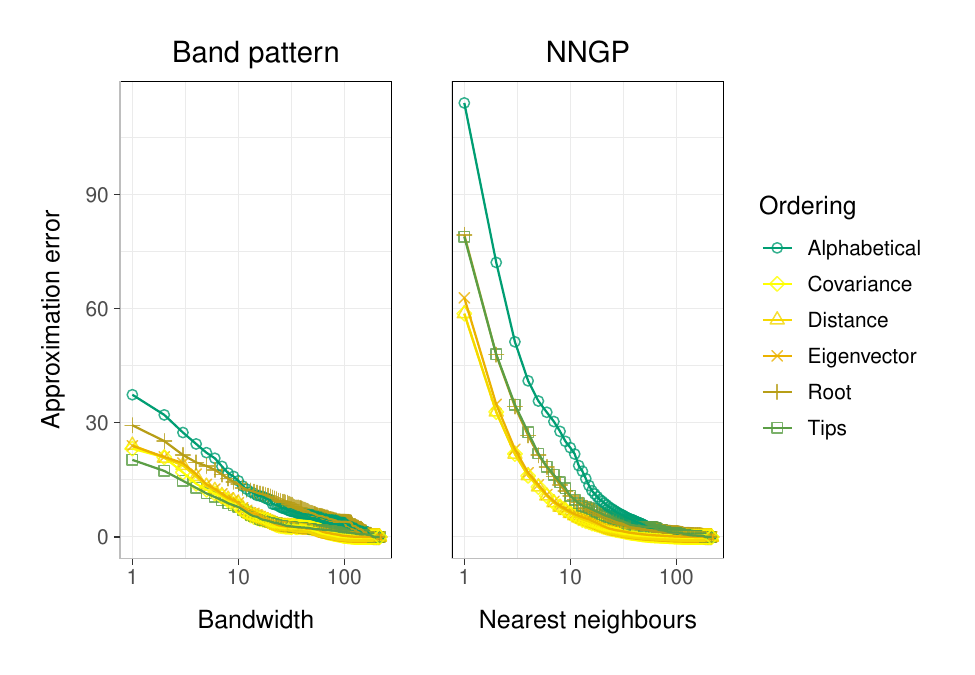}
\caption{Approximation error for the band pattern approximation and the
NNGP, for different permutations of species in the fungi data.}
\end{figure}

The results show that the NNGP has the potential of a smaller
approximation error, though it requires conditioning on six or more
species. When conditioning on less than six species, the pre-defined
band pattern approximation exhibits a smaller approximation error. For
the band pattern approximation, ordering the response data following the
order of species in the phylogeny provides the smallest approximation
error, while for the NNGP this is when ordering the response data by the
total shared branch length in the phylogeny. In both instances, ordering
species alphabetically made for the worst approximation, even though as
a result species within the same genus will usually be located in close
proximity (which can be expected to be phylogenetically similar, and
thus have large covariances). The same does not hold for genera that are
phylogenetically similar.

\begin{figure}
\centering
\includegraphics{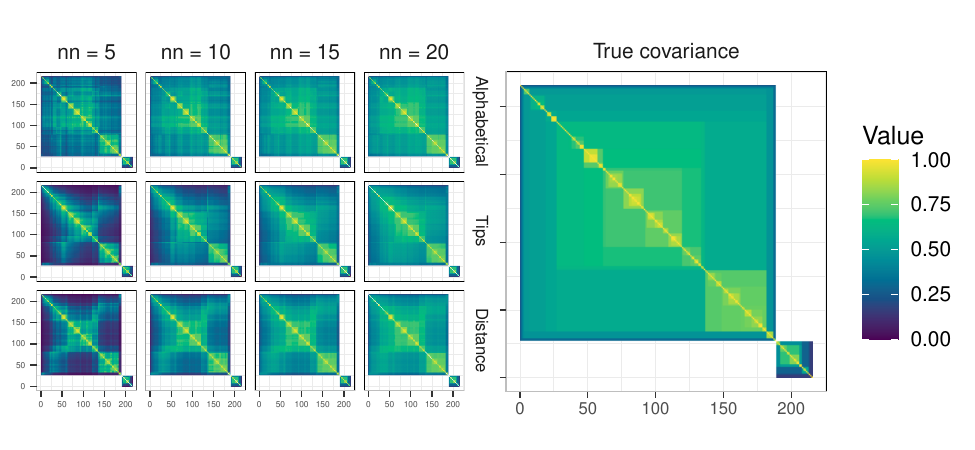}
\caption{The phylogenetic covariance matrix as a result from the sparse
approximation to the precision matrix for different orderings and
numbers of nearest neighbours (nn). Three orderings are displayed,
corresponding to poor (alphabetical), average (tips) and good (distance)
approximations.}
\end{figure}

The approximation error affects both the estimation accuracy and
computation speed. To demonstrate this, we fitted three models to the
example data using three different orderings and
\(nn \in \{5,10,15,20\}\), the results are displayed in Table 1. The
models included three trait covariates, all environmental covariates,
random intercepts for region and reserve. We also assumed a diagonal
\(\textbf{A}^r\) for faster fitting. In Figure 5 the community mean
responses and species' random effects for the model with \(nn = 20\)
with distance ordering are displayed alongside the phylogeny.

\begin{table}[H]
\centering\centering
\caption{\label{tab:restab}Phylogenetic signal parameter estimates for models with different orderings of species in the response data, and number of nearest neighbours (nn), fitted to the Abrego et al. (2022) data. A parameter estimate of 0.05 or lower is displayed as a blank cell for improved readability.}
\centering
\resizebox{\ifdim\width>\linewidth\linewidth\else\width\fi}{!}{
\begin{tabular}[t]{ccccccccccc}
\toprule
Order & nn & Time (min) & Intercept & Deadwood size & Decay stage & Decay stage² & Connectivity & Temperature & Precipitation & ln(Reserve area)\\
\midrule
alphabetical & 5 & 8.77 & 0.13 &  & 1.00 & 0.88 &  & 0.13 & 0.51 & 0.55\\
alphabetical & 10 & 5.92 & 0.24 & 0.65 & 1.00 & 0.87 &  & 0.10 & 0.74 & 0.68\\
alphabetical & 15 & 5.96 & 0.22 & 0.66 & 1.00 & 0.88 &  & 0.14 & 0.71 & 0.70\\
alphabetical & 20 & 9.33 & 0.22 & 0.64 & 1.00 & 0.88 &  & 0.15 & 0.64 & 0.71\\
\cellcolor[HTML]{D3D3D3}{tips} & \cellcolor[HTML]{D3D3D3}{5} & \cellcolor[HTML]{D3D3D3}{11.98} & \cellcolor[HTML]{D3D3D3}{0.17} & \cellcolor[HTML]{D3D3D3}{0.62} & \cellcolor[HTML]{D3D3D3}{0.99} & \cellcolor[HTML]{D3D3D3}{0.89} & \cellcolor[HTML]{D3D3D3}{0.10} & \cellcolor[HTML]{D3D3D3}{} & \cellcolor[HTML]{D3D3D3}{} & \cellcolor[HTML]{D3D3D3}{}\\
\cellcolor[HTML]{D3D3D3}{tips} & \cellcolor[HTML]{D3D3D3}{10} & \cellcolor[HTML]{D3D3D3}{6.96} & \cellcolor[HTML]{D3D3D3}{0.07} & \cellcolor[HTML]{D3D3D3}{} & \cellcolor[HTML]{D3D3D3}{0.98} & \cellcolor[HTML]{D3D3D3}{0.89} & \cellcolor[HTML]{D3D3D3}{0.52} & \cellcolor[HTML]{D3D3D3}{} & \cellcolor[HTML]{D3D3D3}{0.74} & \cellcolor[HTML]{D3D3D3}{}\\
\cellcolor[HTML]{D3D3D3}{tips} & \cellcolor[HTML]{D3D3D3}{15} & \cellcolor[HTML]{D3D3D3}{7.81} & \cellcolor[HTML]{D3D3D3}{0.49} & \cellcolor[HTML]{D3D3D3}{} & \cellcolor[HTML]{D3D3D3}{0.99} & \cellcolor[HTML]{D3D3D3}{0.90} & \cellcolor[HTML]{D3D3D3}{0.59} & \cellcolor[HTML]{D3D3D3}{} & \cellcolor[HTML]{D3D3D3}{0.77} & \cellcolor[HTML]{D3D3D3}{}\\
\cellcolor[HTML]{D3D3D3}{tips} & \cellcolor[HTML]{D3D3D3}{20} & \cellcolor[HTML]{D3D3D3}{8.07} & \cellcolor[HTML]{D3D3D3}{0.50} & \cellcolor[HTML]{D3D3D3}{} & \cellcolor[HTML]{D3D3D3}{0.99} & \cellcolor[HTML]{D3D3D3}{0.91} & \cellcolor[HTML]{D3D3D3}{0.59} & \cellcolor[HTML]{D3D3D3}{} & \cellcolor[HTML]{D3D3D3}{0.81} & \cellcolor[HTML]{D3D3D3}{0.11}\\
distance & 5 & 5.63 & 0.18 &  & 0.68 & 0.88 & 0.12 & 0.10 & 0.16 & \\
distance & 10 & 6.02 & 0.26 &  & 0.99 & 0.86 &  & 0.07 &  & \\
distance & 15 & 8.74 & 0.31 &  & 0.99 & 0.85 &  &  &  & \\
distance & 20 & 6.07 & 0.31 &  & 0.98 & 0.87 & 0.19 &  & 0.46 & \\
\bottomrule
\end{tabular}}
\end{table}

\begin{figure}
\centering
\includegraphics{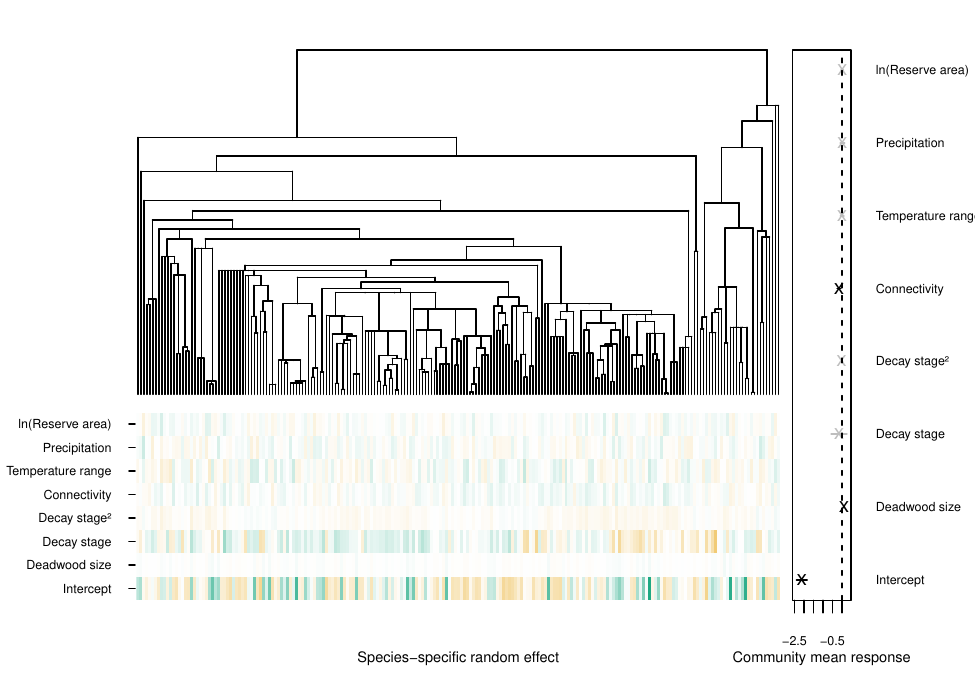}
\caption{The right panel displays community mean responses to the
covariates, with 95\% confidence intervals. Coefficients are greyed out
if the confidence interval crosses zero, so that it is unclear if there
is an effect of a covariate on the community. The bottom panel presents
species-specific deviations from the community mean effects. A
species-specific effect of zero is interpreted as lack of evidence that
a species responds differently to the environment than the community
mean response. Orange colors indicate a more negative response than the
community mean effect, and green more positive. Species effects of which
the prediction interval included zero have been removed.}
\end{figure}

\section{Discussion}\label{discussion}

In this article, we developed a new method for quickly fitting
phylogenetic mixed-effects models, akin to the approaches in the
\texttt{phyr} (Ives et al. 2020), \texttt{Hmsc} (Tikhonov et al. 2024),
and \texttt{MCMCglmm} (Hadfield 2010) \texttt{R}-packages. Because our
models are developed as part of the \texttt{gllvm} \texttt{R}-package
(Niku et al. 2023), the software supports a wide range of response
types, including for Bernoulli, (zero-inflated) Poisson and
negative-binomial, Tweedie, and beta responses (Korhonen et al. 2021;
Niku et al. 2017; Hui et al. 2017).

By utilizing approximate likelihood methods, the models are magnitudes
faster to fit than in the \texttt{Hmsc} and \texttt{MCMCglmm}
\texttt{R}-packages, which are based on Markov chain Monte Carlo
methods. By stacking approximations, namely a VA approximation with
matrix normal covariance structure, combined with low-rank approximation
of one of the matrices, and NNGPs (Finley et al. 2019) for calculating a
sparse approximation to the phylogenetic precision matrix, computation
time of the proposed models is reduced as much as possible. Two sets of
simulation studies confirm that the proposed combination of
approximations still enjoys accurate estimation of the phylogenetic
signal, while significantly speeding up model fitting.

We purposefully limited the flexibility of the VA approximation to fit
phylogenetic random effects models more quickly. The flexibility of VA
is determined by the number of hyperparameters introduced to the
likelihood, which are estimated jointly with the model parameters.
Especially for VA, we find that a good optimisation routine, besides
decent initial values, is critical for successful and speedy convergence
of the models. In our experience, limited-memory optimisation algorithms
(Nocedal and Wright 2006) often lead to faster model fitting and
convergence, and an improved ability to successfully leverage methods
for parallel computation. In future research we hope to explore improved
optimisation routines (as in van der Veen 2024; Kanzow and Steck 2023),
and their ability to more efficiently fit mixed-effects models. In our
simulation study the VA approximation was competitive in computational
speed to the Laplace approximation as applied using Template Model
Builder (Kristensen et al. 2015), though this was limited to 200 species
and five covariates, while we expect the VA approximation to especially
outperform the Laplace approximation with a higher number of covariates
and species. When reducing the number of parameters in the VA
approximation as much as possible, the method fits faster than when
models are fitted using the Laplace approximation.

The cost of quickly fitting the models comes in form of selecting the
number of nearest neighbours for the NNGP approximation. The models are
sensitive to the number of neighbours, which thus needs to be carefully
assess prior to model fitting. Fortunately, this can straightforwardly
be done by quantifying the approximation error under different orderings
of species in the data, and numbers of nearest neighbours, and by
picking the best combination. We demonstrated picking the number of
nearest neighbours and species' ordering using real data on 215 species
of wood-inhabiting fungi (Abrego et al. 2022b). One of the main
novelties of the proposed approach for fitting phylogenetic random
effects models is the ability to estimate phylogenetic signal separately
for each environmental covariate, without the need to introduce an
additional set of random effects as in the \texttt{phyr}
\texttt{R}-package, and while simultaneously facilitating the estimation
of correlation between random effects for different environmental
covariates. Abrego et al. (2022a) fitted various models with and without
trait covariates to determine if phylogenetic structuring of species
responses occurred at multiple spatial scales simultaneously: at the
forest site-scale and at the scale of deadwood logs. Abrego et al.
(2022a) post-hoc estimated covariate-specific phylogenetic signal, while
here we estimated the covariate-specific phylogenetic signal directly as
part of the model, and while fitting the models in minutes, rather than
days as in the \texttt{Hmsc} \texttt{R}-package.

The phylogenetic signal parameter represents the degree to which species
environmental responses are phylogenetically structured, and the
associated variance of covariates the rate at which the responses
evolve. When the phylogenetic signal parameter is one, species responses
to an environmental covariate are fully phylogenetically structured, and
at zero the responses are independent (Pearse, Davies, and Wolkovich
2023). The reason for the presence of phylogenetic signal, or the lack
thereof, cannot be discerned from phylogenetic random effects models
directly. When there is phylogenetic signal, this can be due to the lack
of (phylogenetically structured) continuous trait covariates in the
model, because the measured trait covariates are not phylogentically
structured, or because they are structured in a way that disagrees with
the structuring in species environmental responses. In contrast, a lack
of phylogenetic signal may be because the trait covariates in the model
sufficiently introduce phylogenetic structuring to the species
responses, because there is no ongoing evolution in the trait that is
species environmental responses, or because evolution moves so rapidly
that closely related species are (apparently) so different that their
environmental responses are independent.

Various avenues for future research exist to extend the model. Here, we
assumed the covariance matrix for the covariate effects to be
unstructured. There are various scenarios where a structured covariance
matrix would me more sensible, such as for phylogenetically structured
site intercepts. Sites are not usually randomly distributed in space or
time, while structuring the random effects with an autoregressive or
spatial autocorrelation function (e.g., as in Brooks et al. 2017) could
address such issues. For computational reasons, such an extension would
require an additional approximation to the spatial covariance matrix
(similar to the methods used in this article). Although technically
possible, combining (control of) further approximations with an
intuitive interface might prove challenging.

In their current form phylogenetic random effect models only incorporate
positive species associations, meaning that more closely related species
can only occur in more similar environments. This makes sense if closely
related species segregate their resource utilization. However, two
closely related species that share the same resource would be more
likely to occur in different environments, i.e., exhibit a negative
association. Phylogenetic repulsion can be modelled in the current
framework, by prior to model fitting replacing the phylogenetic
correlation matrix by its inverse (Ives and Helmus 2011). Still, the
current model is limited to either attraction or repulsion, and ideally
its form should be adjusted to allow the data to drive that decision
instead.

\section{Acknowledgements}\label{acknowledgements}

The authors thank Sara Taskinen for comments on an earlier version of
the manuscript.

\section{Data availability}\label{data-availability}

The data used in this study is available online (Abrego et al. 2022b).

\section*{References}\label{references}
\addcontentsline{toc}{section}{References}

\phantomsection\label{refs}

\end{document}